\documentclass [twocolumn,showpacs,preprintnumbers,amsmath,amssymb,aps]{revtex4}
\usepackage {graphicx}
\usepackage {dcolumn}
\usepackage {bm}

\begin {document}


\title {Generalization of Luttinger's Theorem for Strongly Correlated Electron Systems}

\author {M.M. Korshunov}
\email {mkor@iph.krasn.ru}
\author {S.G. Ovchinnikov}
\affiliation {%
L.V. Kirensky Institute of Physics, Siberian Branch of Russian Academy 
of Sciences, 660036 Krasnoyarsk, Russia
}%

\date {\today}

\begin {abstract}

Analyzing general structure of the Green function of a strongly correlated electron system we have shown that for the regime of strong correlations Luttinger's theorem should be generalized in the following way: the volume of the Fermi surface of the system of noninteracting particles is equal to volume of the Fermi surface of the quasiparticles in the strongly correlated system with account for the spectral weight of these quasiparticles. Hubbard and t-J models analysis in the paramagnetic nonsuperconducting phase shows that the generalized Luttinger's theorem is valid for these models.

\end {abstract}

\maketitle

\section {Introduction}

At present, it is widely believed that Luttinger's theorem \cite{ref1} is violated for strongly correlated electron systems (SCES), among which is, in particular, the broad class of superconducting cuprates. This theorem states that the volume of the Fermi surface of interacting particles is equal to that of the noninteracting particles. The proof of Luttinger's theorem \cite{ref1} is valid for the normal Fermi liquid only. A topological non-perturbative proof of this theorem for SCES was given in \cite{ref2} on the assumption that these systems are normal Fermi liquids. Since this proof is based on general considerations it is valid for the t-J model and the Hubbard model in the Fermi liquid phase. However in the SCES other phases are also exist, whose properties differ from those of the Fermi liquid. Deviations from the Fermi liquid behavior reveal themselves in the redistribution of the spectral weight of a quasiparticle between different Hubbard subbands and in the fact that the imaginary part of the self-energy ${\rm Im} \Sigma _k \left( E \right)$ is nonzero on the Fermi surface. Indeed, the relationship ${\rm Im} \Sigma _k \left( E \right) \propto \left( {E - \varepsilon _F } \right)^2 $ is valid in the vicinity of the Fermi level $\varepsilon _F$ of the Fermi liquid.

Calculations of the Hubbard model in the framework of the dynamical mean field theory (DMFT), which is exact in the limit of infinite dimensionality ($d = \infty$) \cite{ref3,ref4,ref5}, demonstrated that the distribution function of the quasiparticles has a jump in the vicinity of the Fermi level. This jump continuously decreases down to zero with increasing the parameter of the on-site Coulomb repulsion $U$. Nevertheless, the Fermi liquid picture persist up to a certain critical value $U_c$ in this case, after which the system transfers to the insulating state. Edwards and Hertz \cite{ref6}, however, using an interpolation scheme for the Hubbard model not based on the limit $d = \infty$ demonstrated that the imaginary part of the self-energy is nonzero on the Fermi surface at $U \propto U_c$ and near the half-filling ($x \ll 1$, where $n = 1 - x$ is the electron concentration). But with a deviation from the half-filling, the Fermi liquid properties of the system are restored fairly rapidly. That the Fermi liquid properties in the Hubbard model with $U = 8t$ ($t$ is the hopping integral) begin to be restored already at $x > 0.1$ was shown in \cite{ref7} in the framework of the non-perturbative approach of dynamical cluster approximation (DCA). At present, the problem of the transition from the Fermi liquid phase to a metallic non-Fermi liquid state with strong electron correlations and the behavior of the system in the transition region has only been stated and is far from being solved (see e.g. \cite{ref8}).

Since in real materials the transition to the metal state is observed at $x \ll 1$, it is the most interesting range of concentrations. At extremely small values of $x$ it would be expected that the additional carriers will be localized in the vicinity of the bottom of the band and the condition ${\rm Im} \Sigma _k \left( {\varepsilon _F } \right) \ne 0$
will be valid for them. As $x$ increases further the Fermi level falls within the range of delocalized states, for which the imaginary part of the self-energy is equal to zero but the non-Fermi liquid effects are still present due to redistribution of the spectral weight between the Hubbard subbands. Deviations from Luttinger's theorem for the Hubbard model in the region where ${\rm Im} \Sigma _k \left( {\varepsilon _F } \right) \ne 0$ were discussed in \cite{ref9} in the framework of the FLEX approximation. In this paper, we restrict ourselves to the concentration range with ${\rm Im} \Sigma _k \left( {\varepsilon _F } \right) = 0$.

\begin {figure}
\includegraphics[width=\linewidth]{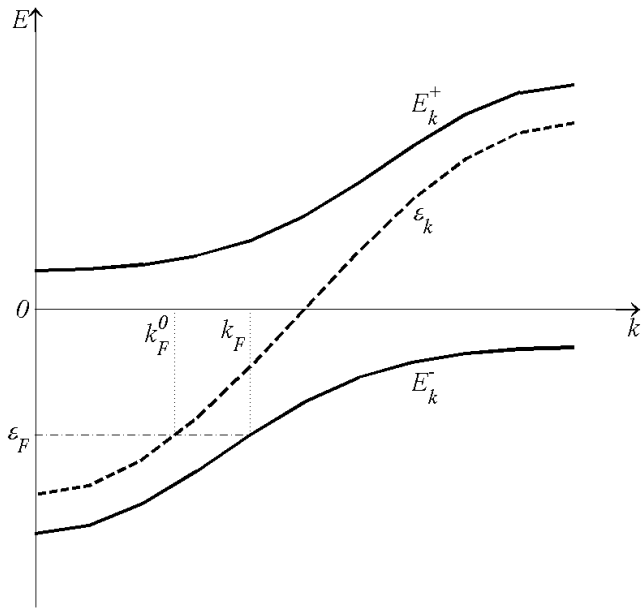}
\caption {\label{fig1} Dispersion curves of the Hubbard bands $E_k^ \pm  $ and the
one-particle spectrum $\varepsilon _k $. $\varepsilon _F $ is the Fermi level, $k_F $ and
$k_F^0 $ are the Fermi momenta for the Hubbard bands and free electrons respectively.}
\end {figure}

As for the case when the spectral weight of a quasiparticle in SCES is not equal to unity, the properties of the system are different from those of the normal Fermi liquid and the original Luttinger theorem is violated. Indeed, the Fermi momentum of the Hubbard bands is larger than the Fermi momentum for free electrons at the same Fermi energy (Fig.~\ref{fig1}), so the geometric volume of the Fermi surface is larger for the Hubbard bands \cite{ref10}. However, in this case the system is in the metallic state and since the distribution function of quasiparticles undergoes a jump in the vicinity of the Fermi level, Luttinger's theorem can be generalized to quasiparticles in the following way: the volume of the Fermi surface of noninteracting particles is equal to that of the interacting quasiparticles with account for the spectral weight of these quasiparticles. In the present work, by analyzing the general structure of the Green function and thorough investigation of the Hubbard-I solution of the t-J mode and Hubbard model \cite{ref11}, we shown that this generalized formulation of Luttinger's theorem is valid for metallic SCES. Actually, such a metallic system is not rather a normal but a ``compressible'' Fermi liquid, which is due to the spectral weight of the quasiparticles being different from unity and to the Fermi surface being less compact (less ``dense''). This ideas makes it possible to eliminate the inconsistency between the concentration of the excess carriers in the superconducting cuprates and the unduly large volume of the Fermi surface calculated in the framework of the model of the normal Fermi liquid.

\section{Analysis of the Green functions' general structure}

Luttinger has shown \cite{ref1} that the equality between the volumes of the Fermi surface in the momentum space for interacting particles $V_{FS}$, and for noninteracting particles $V_{FS}^0 $,
\begin{equation}
\label{eq1}
V_{FS} = V_{FS}^0 ,
\end{equation}
follows from the fact that the average number of particles $\left\langle N \right\rangle $ for interacting and noninteracting fermions is the same. Indeed, for a system without interaction, we have
\begin{eqnarray}
\label{eq2}
\left\langle N \right\rangle &=& \sum\limits_k {\theta \left( {\mu _0 - 
\varepsilon _k } \right)} \nonumber \\
&=& \frac{V}{\left( {2\pi } \right)^3}\int {dk\theta 
\left( {\mu _0 - \varepsilon _k } \right)} = \frac{V}{\left( {2\pi } 
\right)^3}V_{FS}^0 .
\end{eqnarray}
and, for the Fermi liquid system with interaction \cite{ref1}, we have
\begin{eqnarray}
\label{eq3}
\left\langle N \right\rangle &=& \sum\limits_k {\theta \left( {\mu - 
\varepsilon _k - {\rm Re}\Sigma _k } \right)} \nonumber \\
&=& \frac{V}{\left( {2\pi } 
\right)^3}\int {dk\theta \left( {\mu - \varepsilon _k - {\rm Re}\Sigma _k } 
\right)} = \frac{V}{\left( {2\pi } \right)^3}V_{FS} .
\end{eqnarray}
Here, $V$ is the volume of the system of fermions; $\mu $ and $\mu _0 $ are the chemical potentials of the system with and without interaction, respectively; $\varepsilon _k $ are one-particle energies; ${\rm Re}\Sigma _k $ is the self-energy part of the Green function; and $\theta \left( x \right)$ is the Heaviside unit-step function.

For strongly correlated electron systems, however, the definition of the average number of particles as the sum of the Heaviside unit-step functions is invalid, because the spectral weight of each quasiparticle in the system is taken to be the unit in this definition. One of
the essential peculiarities of strongly correlated electron systems is a variation of the spectral weight from unity in each band due to its redistribution between the Hubbard subbands at $U \gg W$ ($W$ is the half-width of the band). For this reason, analogs of Eqs. (\ref{eq2}) and (\ref{eq3}) should be derived for this case.

In what follows, we use the Hubbard X-operators \cite{ref12} defined in the following way:
$X_f^{pq} \equiv \left| p \right\rangle \left\langle q \right|$, where $\left| p \right\rangle $ and $\left| q \right\rangle $ are states at site $f$. Since the number of root vectors $\alpha \left( {p,q} \right)$ is finite, they can be enumerated; thus, we have
\begin{equation}
\label{eq4}
X_f^{pq} \leftrightarrow X_f^{\alpha \left( {p,q} \right)} \leftrightarrow 
X_f^{\alpha _m } \leftrightarrow X_f^m .
\end{equation}
Here, index $m \leftrightarrow \left( {p,q} \right)$ enumerates quasiparticles
with the energies
\begin{equation}
\label{eq5}
\omega _m = \omega _{pq} = \varepsilon _p \left( {N + 1} \right) - 
\varepsilon _q \left( N \right),
\end{equation}
where $\varepsilon _p $ is the energy level with index $p$ for the $N$-electron system.

The Hubbard operators are related to one-electron creation and annihilation operators in the following way:
\begin{equation}
\label{eq6}
a_{f\lambda \sigma }^ + = \sum\limits_m {\gamma _{\lambda \sigma }^* \left( m \right)\mathop {X_{f\sigma }^m }\limits^ + },
a_{f\lambda \sigma } = \sum\limits_m {\gamma _{\lambda \sigma } \left( m \right)X_{f\sigma }^m },
\end{equation}
where $\gamma _{\lambda \sigma } \left( m \right)$ is the partial weight of a quasiparticle $m$ with spin $\sigma $ and orbital index $\lambda $.

The average occupation numbers $\left\langle {n_{k\lambda \sigma } } \right\rangle $ for the particles with momentum $k$ and spin $\sigma$ are expressed in terms of the electron Green function written in the energy representation, $G_{k\lambda \sigma } = \left\langle {\left\langle {a_{k\lambda \sigma } } \mathrel{\left| {\vphantom {{a_{k\lambda \sigma } } 
{a_{k\lambda \sigma }^ + }}} \right. \kern-\nulldelimiterspace} {a_{k\lambda \sigma }^ + } \right\rangle } \right\rangle _{E + i\eta }$, in the following way:
\begin{equation}
\label{eq7}
\left\langle {n_{k\lambda \sigma } } \right\rangle = \int {dE\,f_F \left( E 
\right)\,\left( { - \frac{1}{\pi }{\rm Im}G_{k\lambda \sigma } } \right)} ,
\end{equation}
where $f_F(E)$ is the Fermi function, $\eta \to 0$, $\eta > 0$.
In the X-operators representation, the Green function has the form
\begin{eqnarray}
\label{eq8}
&\left\langle {\left\langle {a_{k\lambda \sigma } } \mathrel{\left| 
{\vphantom {{a_{k\lambda \sigma } } {a_{k\lambda \sigma }^ + }}} \right. 
\kern-\nulldelimiterspace} {a_{k\lambda \sigma }^ + } \right\rangle } 
\right\rangle _{E + i\eta } \nonumber& \\ 
&=\sum\limits_{m,p} {\gamma _{\lambda \sigma } 
\left( m \right)\gamma _{\lambda \sigma }^\ast \left( p \right)\left\langle 
{\left\langle {X_{k\sigma }^m } \mathrel{\left| {\vphantom {{X_{k\sigma }^m 
} {\mathop {X_{k\sigma }^p }\limits^ + }}} \right. 
\kern-\nulldelimiterspace} {\mathop {X_{k\sigma }^p }\limits^ + } 
\right\rangle } \right\rangle _{E + i\eta } }.&
\end{eqnarray}

For the matrix Green function $D_{k\sigma }^{m,p} \left( E \right) = \left\langle {\left\langle {X_{k\sigma }^m } \mathrel{\left| {\vphantom {{X_{k\sigma }^m } {\mathop {X_{k\sigma }^p }\limits^ + }}} \right. \kern-\nulldelimiterspace} {\mathop {X_{k\sigma }^p }\limits^ + } 
\right\rangle } \right\rangle _{E + i\eta } $, the generalized Dyson equation \cite{ref13} can be written as
\begin{equation}
\label{eq9}
\hat {D}_{k\sigma } \left( E \right) = \left\{ {\left[ {\hat {G}_{k\sigma 
}^{(0)} \left( E \right)} \right]^{ - 1} + \hat {\Sigma }_{k\sigma } \left( 
E \right)} \right\}^{ - 1}\hat {P}_{k\sigma } \left( E \right).
\end{equation}

Here, $\hat {\Sigma }_{k\sigma } \left( E \right)$ and $\hat {P}_{k\sigma } \left( E \right)$ are the self-energy and the force operator, respectively. The presence of the force operator is due to the redistribution of the spectral weight and is an intrinsic feature of SCES. The concept of the force operator was introduced earlier in a diagram technique for spin systems
\cite{ref14}. The Green function $\hat {G}_{k\sigma }^{(0)} \left( E \right)$ in Eq. (\ref{eq9}) is defined by the formula
\begin{equation}
\label{eq10}
\left[ {\hat {G}_{k\sigma }^{(0)} \left( E \right)} \right]^{ - 1} = \hat 
{G}_0^{ - 1} \left( E \right) - \hat {P}_{k\sigma } \left( E \right)\hat 
{T}_{k\sigma } ,
\end{equation}
where $\hat {T}_{k\sigma } $ is the interaction matrix element (for the Hubbard model, $T_{k\sigma }^{m,p} = \gamma _\sigma \left( m \right)\gamma _\sigma 
^\ast \left( p \right)t_k$).

In the Hubbard-I approximation at $U \gg W$, the structure of the exact Green function (\ref{eq9}) remains unchanged but the self-energy is equal to zero and the force operator $P_{k\sigma }^{m,p} \left( E \right)$ is replaced by $P_{k\sigma }^{m,p} \left( E \right) \to P_{0\sigma }^{m,p} = \delta _{m,p} F_\sigma ^m $, where $F_\sigma ^m \equiv F\left( {p,q} \right) = \left\langle {X_f^{pp} } \right\rangle + \left\langle {X_f^{qq} } \right\rangle $ is the occupation factor, which is referred to as the end factor in the diagram technique for the X operators developed in \cite{ref15}. In terms of the Hubbard-I approximation, the following formula is derived from Eq. (\ref{eq9}):
\begin{equation}
\label{eq11}
\hat {D}_{k\sigma }^{(0)} = \left\{ {\hat {G}_0^{ - 1} \left( E \right) - 
\hat {P}_{0\sigma } \hat {T}_{k\sigma } } \right\}^{ - 1}\hat {P}_{0\sigma }.
\end{equation}

In order to estimate the contributions to Eq. (\ref{eq11}) in higher order approximations (with respect to the Hubbard-I approximation), let us compare the exact equation (\ref{eq9}) for the Green function with Eq. (\ref{eq11}), written in the Hubbard-I approximation. First, there is a difference due to the renormalization of the occupation factors $F_\sigma ^m $ which arises when the exact equation for the force operator $\hat {P}_{k\sigma } \left( E \right)$ is used. However, taking into account the corrections due to the force operator keeps the structure of the Hubbard bands unchanged and, therefore, does not lead to a qualitative difference of the exact Green function from that in the Hubbard-I approximation. A second essential difference is the renormalization of the real part of the self-energy $\hat {\Sigma }_{k\sigma } \left( E \right)$ and the appearance of quasiparticle damping. The latter implies non-Fermi liquid behavior of the system and, as mentioned above, the consideration of the region where ${\rm Im}\Sigma _k \left( {\varepsilon _F } \right) = 0$. In this region, the exact Green function given by Eq. (\ref{eq9}) can be rewritten as the sum of one-pole contributions over the quasiparticle bands labeled by index $\xi $ (for the t–J model, $\xi $ has one value, $\xi = 1$; for the Hubbard model, $\xi = 1,2$). In the general multiband case, the exact Green function is
\begin{equation}
\label{eq12}
G_{k\lambda \sigma } \left( E \right) = \sum\limits_\xi {\frac{F_{k\lambda 
\sigma } \left( \xi \right)}{E - \Omega _{k\sigma } \left( \xi \right) + \mu 
+ i\eta }}.
\end{equation}
Here, the real part of the self-energy contributes not only to the renormalization of the dispersion law but also to the renormalization of the spectral weight. Such a representation for the electron Green function has been obtained earlier in the Hubbard model in terms of
the spectral density approach (SDA) \cite{ref16}. This approach is non-perturbative and assumes only the absence of the quasiparticle damping. The spectral weights $F_{k\lambda \sigma } \left( \xi \right)$ and the band energies $\Omega _{k\sigma } \left( \xi \right)$ are calculated in the SDA by using the method of moments (see the review and comparison with other methods in \cite{ref17}).

As for the renormalization of the real part of the self-energy, this effect introduces corrections to the energy spectrum $\Omega _{k\sigma } \left( \xi \right)$ and qualitatively does not change the further reasoning. The fact that the structure of the Green function is correct even in the Hubbard-I approximation (it is the structure of the Green function that is essential for further derivation of Luttinger's theorem) follows from a comparison of the Hubbard-I solution and the exact solution in the infinite-dimensionality limit obtained in DMFT \cite{ref4,ref5}, as well as from a comparison of the Hubbard-I solution and a numerical solution obtained using the exact Quantum Monte Carlo (QMC) method for the Hubbard model \cite{ref18,refAlena}. A comparison of the spectral densities obtained in the Hubbard-I approximation at $U \gg W$ and those derived by the QMC shows numerical coincidence between them as in paramagnetic \cite{ref18} as in antiferromagnetic phases \cite{refAlena}. In terms of the diagram technique for the X-operators, it has also been demonstrated that this approximation gives simple and pictorial relationships which correctly describe the physics of the phenomena at $U \gg t$ \cite{ref13,ref19}.

Substituting Eq. (\ref{eq12}) into Eq. (\ref{eq7}) and using the spectral theorem, we obtain
\begin{eqnarray}
\label{eq13}
\left\langle {n_{k\lambda \sigma } } \right\rangle &=& \int {dE\,f_F \left( E 
\right)\sum\limits_\xi {F_{k\lambda \sigma } \left( \xi \right)\delta 
\left( {E - \Omega _{k\sigma } \left( \xi \right) + \mu } \right)} } \nonumber \\
&=& \sum\limits_\xi {F_{k\lambda \sigma } \left( \xi \right)f_F \left( {\Omega 
_{k\sigma } \left( \xi \right) - \mu } \right)} ,
\end{eqnarray}
Taking into account that the quantities in Eq. (\ref{eq13}) do not depend on spin in the paramagnetic phase, the average number of particles $\left\langle N \right\rangle $ at zero temperature can be written in the compact form
\begin{equation}
\label{eq14}
\left\langle N \right\rangle = \sum\limits_{k,\lambda } {\left\langle 
{n_{k\lambda \sigma } } \right\rangle } = \sum\limits_{k,\xi } {F_k \left( 
\xi \right)\theta \left( {\mu - \Omega _k \left( \xi \right)} \right)} ,
\end{equation}
where $F_k \left( \xi \right) = 2\sum\limits_\lambda {F_{k\lambda \sigma } \left( \xi \right)}$.

For noninteracting particles, we have $F_k \left( \xi \right) = 1$ and the equation for $\left\langle N \right\rangle $ 
\begin{eqnarray}
\label{eq15}
\left\langle N \right\rangle &=& \sum\limits_k {\theta \left( {\mu _0 - 
\varepsilon _k } \right)} = \frac{V}{\left( {2\pi } \right)^3}\int {dk\theta 
\left( {\mu _0 - \varepsilon _k } \right)} \nonumber \\
&=& \frac{V}{\left( {2\pi } \right)^3}V_{FS}^0,
\end{eqnarray}
completely coincides with Eq. (\ref{eq2}) in this case. 

For the system of interacting quasiparticles, Eq. (\ref{eq14}) can be written as
\begin{eqnarray}
\label{eq16}
\left\langle N \right\rangle &=& \frac{V}{\left( {2\pi } 
\right)^3}\sum\limits_\xi {\int {dkF_k \left( \xi \right)\theta \left( 
{\mu - \Omega _k \left( \xi \right)} \right)} } \nonumber \\
&=& \frac{V}{\left( {2\pi } \right)^3}\sum\limits_\xi {V_{FS}^\xi }.
\end{eqnarray}
By comparing Eq. (\ref{eq15}) with Eq. (\ref{eq16}), we obtain
\begin{equation}
\label{eq17}
V_{FS}^0 = \sum\limits_\xi {V_{FS}^\xi } ,
\end{equation}
where $V_{FS}^\xi $ is the volume of the energy subband $\xi $ taking
into account the spectral weight $F_k \left( \xi \right)$ of this subband,
\begin{equation}
\label{eq18}
V_{FS}^\xi = \int {dkF_k \left( \xi \right)\theta \left( {\mu - \Omega _k \left( \xi \right)} \right)}.
\end{equation}

Equation (\ref{eq17}) is the generalized Luttinger theorem: the right-hand side of the equality is a superposition of the volumes $V_{FS}^\xi $ for the different energy subbands $\xi $ rather than the volume $V_{FS} $ and each state $\left| {k,\sigma } \right\rangle $ for the band $\xi$ enters with a decreased spectral weight.

Therefore, the region bounded by the Fermi surface in the $k$ space becomes ``less dense''. Indeed, let us use the relationship $m_{FS} = \rho_{FS} V_{FS} $, where $m_{FS} $ is the ``mass'' of the Fermi surface, $\rho _{FS} $ is its ``density'', and $V_{FS} $ is its volume. It is obvious that the ``mass'' $m_{FS} $ is proportional to the average number of particles $\left\langle N \right\rangle $ and $\rho _{FS} $ is the spectral weight of the quasiparticles under Fermi surface. For the system of noninteracting particles, we have $m_{FS}^0 = \rho_{FS}^0 V_{FS}^0 $. Further, from the equality $m_{FS} = m_{FS}^0 $, it follows that $\rho _{FS} V_{FS} = \rho _{FS}^0 V_{FS}^0 $ and
\begin{equation}
\label{eq19}
V_{FS} = \frac{1}{\rho _{FS} }V_{FS}^0 ,
\end{equation}
since for the system without interaction we have $\rho _{FS}^0 = 1$. It is seen that, if the spectral weight of the quasiparticles differs from unity $V_{FS}$ is not equal to $V_{FS}^0$. It is precisely this case ($\rho _{FS} \le 1$) that is realized in SCES. On the other hand, the quantity given by Eq. (\ref{eq19}) is invariant under the interaction in the system; therefore, the generalization of Luttinger's theorem for quasiparticles is as follows: the volume of the Fermi surface of a system of noninteracting particles is equal to that of interacting quasiparticles with account for the spectral weight of these quasiparticles. This formulation of the theorem is valid for both the band electrons and the quasiparticles in metallic SCES in the limit $U \gg W$.

The deviation of the spectral weight from unity can be considered as the transition to a space with a different metric. This is demonstrated in Section~\ref{section:metric}, in which it is shown that the quantity given by Eq. (\ref{eq19}) rather than by Eq. (\ref{eq1}) is invariant under this transition.

\section{t-J model}

The Hamiltonian of the Hubbard model in the X-operator representation \cite{ref11} has the form
\begin{eqnarray}
\label{eq20}
H &=& \sum\limits_{f,\sigma } {\left( {\left( {\varepsilon - \mu } 
\right)\left( {X_f^{\sigma \sigma } + X_f^{SS} } \right) + 
\frac{U}{2}X_f^{SS} } \right)} \nonumber \\
&+& \sum\limits_{f \ne g,\sigma } {t_{fg} \left( {X_f^{\sigma 0}+2\sigma X_f^{S\bar {\sigma }}}
\right)\left( {X_g^{0\sigma } + 2\sigma X_g^{\bar {\sigma }S} } \right)}.
\end{eqnarray}

The Hamiltonian of the t–J model can be derived form Eq. (\ref{eq20}) in the limit of the strong Coulomb interaction $U \gg t$:
\begin{eqnarray}
\label{eq21}
H_{t - J} &=& \sum\limits_{f,\sigma } {\left( {\varepsilon - \mu } \right)X_f^{\sigma \sigma } } + \sum\limits_{f \ne g,\sigma } {t_{fg} X_f^{\sigma 0} X_g^{0\sigma } } \nonumber \\
&+& J\sum\limits_{f \ne g} {\left( {{\rm {\bf S}}_f {\rm {\bf S}}_g - \frac{1}{4}n_f n_g } \right)},
\end{eqnarray}
where $t_{fg} $ is the hopping integral, $J$ is the exchange integral, ${\rm {\bf S}}_f $ is the spin operator, and $n_f $ is the number of particles operator. Since only one fermi-type root vector $\left\{{X_f^{0\sigma } } \right\} \leftrightarrow \left\{ {X_f^1 } \right\}$ is present then the Green function in the case of ${\rm Im}\Sigma _k \left( E \right) = 0$ has the form
\begin{equation}
\label{eq22}
D_{k\sigma } \left( E \right) = \frac{F\left( 1 \right)}{E - E_k + \mu },
\end{equation}
where $F\left( 1 \right) / 2 = \left( {1 - x} \right) / 2$ is the spectral weight of the only band $\xi = 1$ and $E_k $ is the spectrum of the system in the Hubbard-I approximation:
\begin{equation}
\label{eq23}
E_k = \varepsilon + t_k \left( {\frac{1 - x}{2}} \right) - \frac{J}{2}\left({\frac{1 + x}{2}} \right).
\end{equation}
Here, $t_k = 2t\left( {\cos k_x + \cos k_y } \right)$ is the Fourier transform of
the hopping integral in the case of a square lattice. The number of particles is
\begin{equation}
\label{eq24}
\left\langle N \right\rangle = \sum\limits_{k,\sigma } {\left\langle 
{X_k^{\sigma 0} X_k^{0\sigma } } \right\rangle } = \sum\limits_k {\left( {1 
- x} \right)f_F \left( {E_k - \mu } \right)}.
\end{equation}

At zero temperature, we have
\begin{equation}
\label{eq25}
\left\langle N \right\rangle = \sum\limits_k {\left( {1 - x} \right)\theta \left( {\mu - E_k } \right)} .
\end{equation}

This equation coincides with Eq. (\ref{eq14}). Thus for the volume of the Fermi surface we have $V_{FS}^0 = F\left( 1 \right)V^ - = \left( {1 - x} \right)V^-$, where $V^ - = \int {dk\theta \left( {\mu - E_k } \right)} $. In multiband models, such as the Hubbard model, the spectral weight of the quasiparticles is redistributed between the bands due to strong correlations. In t-J model case, there is only one band, but its spectral weight is smaller than unity [see Eq. (\ref{eq25})] because the part of the spectral weight goes to the upper Hubbard band. In deriving the equations for the t–J model from the Hubbard model, this band is taken into account only in terms of the perturbation theory with respect to the parameter $t / U \ll 1$ and does not appear in the Hamiltonian (\ref{eq21}) because of the constraint of no two particle excitations $\left\langle {X_f^{SS} } \right\rangle \to 0$.

\section{Hubbard model}

The basis fermion operators for the Hubbard model (\ref{eq20}) are $\left\{ {X_f^{0\sigma } ,X_f^{\bar {\sigma }S} } \right\} \leftrightarrow \left\{ {X_f^1 ,X_f^2 } \right\}$, where $\left| S \right\rangle$ is a two-particle singlet, $\left| 0 \right\rangle $ is the vacuum state, and $\left| \sigma \right\rangle $, $\left| \bar {\sigma } \right\rangle $ are one-particle states. The matrix Green function has the form
\begin{widetext}
\begin{equation}
\label{eq26}
\hat {D}_{k\sigma } \left( E \right) = \frac{1}{\det }
\left({{\begin{array}{cc}
 {\frac{F\left( 1 \right)}{2}\left( {E - \varepsilon + \mu - U - t_k \frac{1 + x}{2}} \right)} &
 {2 \sigma t_k \frac{F(2)}{2} \frac{F(1)}{2}} \\
 {2 \sigma t_k \frac{F(1)}{2} \frac{F(2)}{2}} &
 {\frac{F\left( 2 \right)}{2}\left( {E - \varepsilon + \mu - t_k \frac{1 - x}{2}} \right)} 
\end{array}}} \right),
\end{equation}
\end{widetext}
where $\det  = \left( {E - E_k^ + + \mu } \right)\left( {E - E_k^ - + \mu } \right)$, $F\left( 1 \right) / 2 = \left( {1 - x} \right) / 2$ and $F\left( 2 \right) / 2 = \left( {1 + x} \right) / 2$ are the occupation factors of the lower and upper Hubbard bands, respectively. The energy spectrum of the system is
\begin{equation}
\label{eq27}
E_k^\pm = \frac{1}{2}\left( {t_k + U \pm \sqrt {t_k ^2 + U^2 + 2t_k Ux} } \right).
\end{equation}

The number of particles can be easily found using the Green function (\ref{eq26}):
\begin{eqnarray}
\label{eq28}
\left\langle N \right\rangle &=& \sum\limits_k \left\{ \left[ 1 + 
\frac{x\left( U + t_k x \right)}{E_k^ + - E_k^ - } \right]f_F \left( 
E_k^ + - \mu \right) \right. \nonumber \\
&+& \left. \left[ 1 - \frac{x\left( U + t_k x 
\right)}{E_k^ + - E_k^ - } \right]f_F \left( E_k^ - - \mu \right\} 
\right).
\end{eqnarray}
This equation coincides with Eq. (\ref{eq14}). The expressions in square brackets are the spectral weights of the upper and lower Hubbard bands, respectively. Their sum (taking into account the spin) is equal to the spectral weight of noninteracting electrons, as it must be. Now, let us analyze Eq. (\ref{eq28}) in the regime of strong Coulomb repulsion $U \gg t$. In this case, the denominator $E_k^ + - E_k^ - $ can be expanded in powers of the small parameter $t / U \ll 1$. Neglecting second-order terms, we have
\begin{equation}
\label{eq29}
\left\langle N \right\rangle = \sum\limits_k {\left( {\left( {1 + x} 
\right)f_F \left( {E_k^ + - \mu } \right) + \left( {1 - x} \right)f_F \left( 
{E_k^ - - \mu } \right)} \right)}.
\end{equation}
At zero temperature, this equation becomes
\begin{equation}
\label{eq30}
\left\langle N \right\rangle = \sum\limits_k {\left( {\left( {1 + x} 
\right)\theta \left( {\mu - E_k^ + } \right) + \left( {1 - x} \right)\theta 
\left( {\mu - E_k^ - } \right)} \right)} .
\end{equation}

The expression for the volumes of the Fermi surfaces is
\begin{eqnarray}
\label{eq31}
V_{FS}^0 &=& F\left( 2 \right)V^ + + F\left( 1 \right)V^ - \nonumber \\ 
&=& \left( {1 + x} \right)V^ + + \left( {1 - x} \right)V^ -.
\end{eqnarray}
where $V^\pm = \int {dk\theta \left( {\mu - E_k^\pm } \right)} $. It is clearly seen that the due to strong Coulomb interaction the spectral weight is redistributed between the lower and the upper Hubbard subbands. It is this effect that makes generalization of Luttinger's
theorem for strongly correlated electron systems is of essence.

The splitting into two Hubbard bands is clearly seen in Eqs. (\ref{eq29}), (\ref{eq30}) and (\ref{eq31}). It is easy to transform the Hubbard model to the t–J model simply by neglecting the influence of the upper (or lower) band, because the bands are separated by a gap $U$ (the interband hopping was already eliminated by the expansion in powers of $t/U$). The occupation numbers are immediately found to be
\begin{equation}
\label{eq32}
\mathop {\lim }\limits_{U \to \infty } \left\langle N \right\rangle = \sum\limits_k {\left( {1 - x} \right)f_F \left( {E_k^ - - \mu } \right)}.
\end{equation}
This expression coincides with Eq. (\ref{eq24}), obtained for the number of particles in the t–J model. Thus, the decrease in the spectral weight in the t–J model is a result of the approximations used ($t / U \ll 1$) rather than of its spontaneous disappearance.

\begin {figure}
\includegraphics[width=\linewidth]{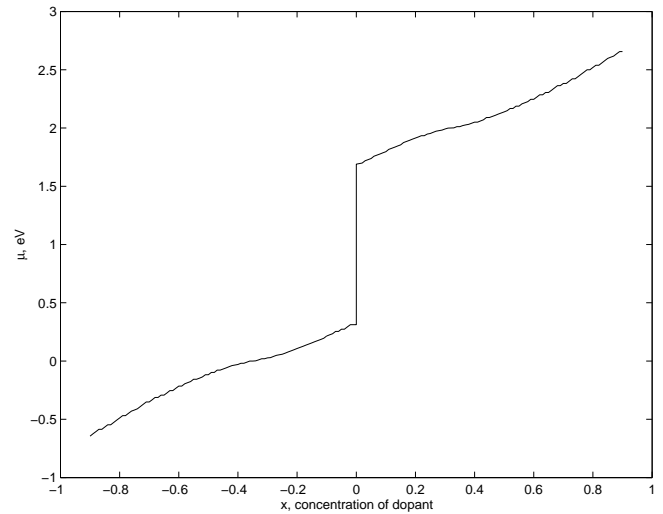}
\caption {\label {fig2} Dependence of the chemical potential $\mu$ on $x$.}
\end {figure}

\begin{figure}
\includegraphics[width=\linewidth]{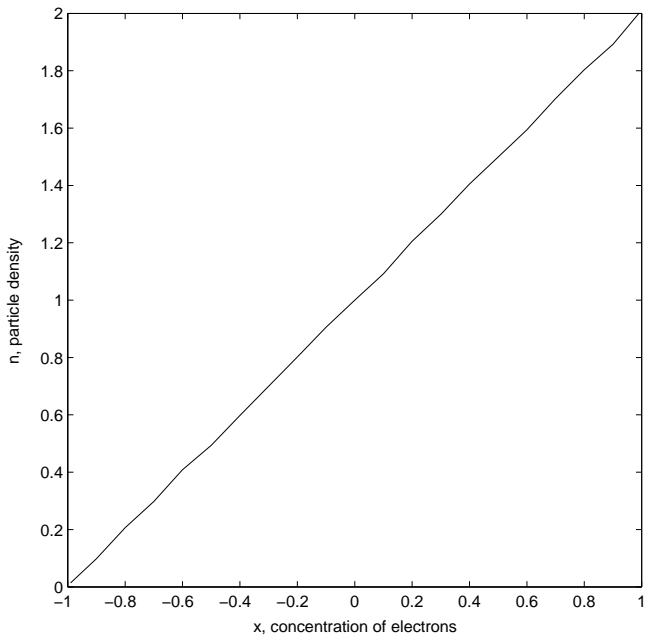}
\caption{\label{fig3} Dependence of the density of particles $n = \left\langle N \right\rangle/N$ on
$x$ ($N$ is the number of vectors in the momentum space).}
\end{figure}

In Figs.~\ref{fig2} and ~\ref{fig3} the results of calculations for zero temperature at $U = 10\vert t\vert $, $t = - 0.2\,eV$ are shown. Calculations for finite temperatures were also carried out, but they did not reveal any qualitative difference from the case of zero temperature. The chemical potential $\mu $ calculated self-consistently by using Eq. (\ref{eq28}) is shown in Fig.~\ref{fig2}. Fig.~\ref{fig3} shows the dependence of the density of particles $n = \left\langle N \right\rangle / N$ on $x$ calculated by using Eq. (\ref{eq29}). It is clearly seen that this dependence is linear and, moreover, $n = 1 + x$. Actually, the last equality means that the generalized Luttinger theorem is valid. Indeed, the left hand side of Eq. (\ref{eq29}) is the number of particles calculated with account for the interaction in the system, while the right hand side of the equation is the number of noninteracting particles. The equality of these two quantities is a prerequisite for the equality of the Fermi surface volumes multiplied by the corresponding spectral weights (\ref{eq31}).

\section{Quasiparticle description as a transformation of the metric space \label{section:metric}}

Let ${\rm {\bf e}}_\mu $ be natural reference vectors associated with the system of curvilinear (in general) coordinates $x^\mu $. In what follows, upper indices indicate contravariant quantities and lower indices, covariant ones. The metric tensor is defined as
\begin{equation}
\label{eq33}
g_{\mu \nu } = \left( {{\rm {\bf e}}_\mu \cdot {\rm {\bf e}}_\nu } \right).
\end{equation}
In going to new coordinates $y^\mu $ we have
\[
{\rm {\bf e}}_\mu = a_\mu ^\nu {\rm {\bf e}}'_\nu ,
\]
\[
g_{\mu \nu } = \left( {{\rm {\bf e}}_\mu \cdot {\rm {\bf e}}_\nu } \right) = 
a_\mu ^\rho a_\nu ^\eta \left( {{\rm {\bf e}}_\rho \cdot {\rm {\bf e}}_\eta 
} \right) = a_\mu ^\rho a_\nu ^\eta \,g'_{\rho \eta } ,
\]
where $a_\mu ^\nu = \frac{\partial y^\nu }{\partial x^\mu }$ are the coefficients of the axis transformation.

By definition, an elementary volume of the $n$-dimensional space is
\begin{equation}
\label{eq34}
d\tau = dx^1 \cdot dx^2 \cdot ... \cdot dx^n.
\end{equation}
In this case, the value $\sqrt { - g} \cdot d\tau $ rather than the volume element $d\tau $ is invariant under transition to another coordinate system. Here, $g = \det g_{\mu \nu } $ is the determinant of the components of the metric tensor, i.e.
\begin{equation}
\label{eq35}
\sqrt { - g'} \cdot d\tau ' = \sqrt { - g} \cdot d\tau .
\end{equation}

Now, let us consider two $n$-dimensional spaces: one for quasiparticles with the spectral weight $\rho '$ (quantities referred to this space are labeled by prime) and one for quasiparticles with the spectral weight $\rho $. Obviously, a transition between these coordinate systems can be made by simply changing the axis scales, $a_\mu ^\nu = \sqrt[n]{\rho \mathord{\left/ {\vphantom {\rho {\rho '}}} \right. \kern-\nulldelimiterspace} {\rho '}}$. The corresponding transformation of the metric tensor is
\begin{equation}
\label{eq36}
g_{\mu \nu } = \left( {\frac{\rho }{\rho '}} \right)^{2 / n} \cdot g'_{\mu 
\nu } .
\end{equation}
The relation between the elementary volumes derived from Eqs. (\ref{eq35}) and (\ref{eq36}) is
\begin{equation}
\label{eq37}
d\tau = \frac{\rho '}{\rho }d\tau '.
\end{equation}
This equation (with $\rho ' = 1$) coincides with Eq. (\ref{eq19}). A similar relationship takes place in the hydrodynamic theory for a compressible liquid. Thus, in the case of quasiparticles with a spectral weight smaller than unity, we deal with modification of the normal Fermi liquid instead of the normal Fermi liquid; by analogy with hydrodynamics, this modification can be called the ``compressible Fermi liquid''. The original Luttinger theorem is valid only for quasiparticles with a spectral weight equal to unity and, therefore, is of limited use. In systems with different spectral weights of quasiparticles, the quantity given by Eq. (\ref{eq35}) rather than the volume of the Fermi surface is conserved; thus, the scalar density of noninteracting particles $\sqrt { - g'} \cdot d\tau '$ is equal to the scalar density of quasiparticles with interaction $\sqrt {- g} \cdot d\tau $. It is clearly seen that Luttinger's theorem \cite{ref1} is a special case of this statement.

\section{Conclusion}

One of the problems of the theory of SCES is whether or not Luttinger's theorem is valid for these systems. This question is of great importance in describing high-temperature superconductors, because they belong to the class of SCES. It has been shown \cite{ref9,ref10,ref18} that in the Hubbard model Luttinger's theorem \cite{ref1} is violated for underdoped samples ($x < x_{opt}$) because due to short-range magnetic order and the spin fluctuations associated with it. However, Luttinger's theorem is valid in the overdoped regime ($x > x_{opt}$), where the paramagnetic metal state takes place. Actually, the original Luttinger theorem in the form of Eq. (\ref{eq1}) is not valid for SCES; the proof of theorem (\ref{eq1}) is not applicable to these systems because it does not take into account the difference of the spectral weight of quasiparticles from unity, which is one of the most remarkable feature of SCES \cite{ref13}. In present paper we have formulated Luttinger's theorem (\ref{eq17}) generalized for the case of quasiparticle description. Qualitative analysis of this generalization given by Eq. (\ref{eq19}), as well as analytically exact derivation of Eq. (\ref{eq37}) for the scalar densities, showed that the region of $k$ space under the Fermi surface becomes less compact (or, in other words, less ``dense'') in quasiparticle systems: the contribution of each state is renormalized by decreased spectral weight of the corresponding quasiparticle. 

The momentum space is divided into quantum cells, each of which can contain one electron or, taking into account the Pauli principle, two electrons with opposite spins. Some states from the whole set of quantum states in a cell move away to infinite energies due to strong electron correlations. Therefore, the spectral weight $F$ of the remaining states is smaller than unity; quasiparticle excitations in such a system become renormalized, and their spectral weight $F < 1$. It is this effect that causes the $k$ space to be less compact. 

It is seen from Eq. (\ref{eq37}) for the scalar densities that we deal with a compressible Fermi liquid. A normal Fermi liquid belongs to a subclass of the class of compressible Fermi liquids; in this subclass, the spectral weights of the quasiparticles are equal to unity. The generalized Luttinger's theorem is formulated for the case of a compressible Fermi liquid in which the effects of strong electron correlations necessitate deviation from the description of the system as a normal Fermi liquid.

In this paper we have considered basic models of strongly correlated systems, such as the t-J model and the Hubbard model. It has been shown that in the nonsuperconducting paramagnetic phase these models satisfy a generalized Luttinger's theorem. In the Hubbard model the spectral weight is redistributed between the Hubbard subbands; in the t-J model a decrease in the spectral weight occurs because part of the states moves away to infinite energies due to the strong correlation between the electrons (the upper Hubbard subband is separated form the lower band by a gap $U \ll t$).

\begin {acknowledgments}
The authors are grateful to V.V. Val'kov for helpful remarks and A. Pushnov for partial translation.
This work was supported by the INTAS grant 01-0654, the Russian Foundation for Basic Research and the Krasnoyarsk Science Foundation program ``Enisey'' 02-02-97705), the program of the Russian Academy of Sciences ``Quantum Macrophysics'', and Siberian Branch of RAS (Lavrent'yev Contest for Young Scientists).
\end {acknowledgments}

\end {document}